\documentclass[aps,prl,twocolumn,showpacs,superscriptaddress,groupedaddress]{revtex4} 

\usepackage{graphicx}  % needed for figures
\usepackage{dcolumn}   % needed for some tables
\usepackage{amssymb}   % for math
\usepackage{soul}
\usepackage{color}
\usepackage{natbib,hyperref}
\usepackage{multirow}

\usepackage{hyperref}

\begin{document}
%\linenumbers
\preprint{DENEMEEEEEEEEEEEE}

\title{
Discovery of $\tau$ neutrino appearance in the CNGS neutrino beam with the OPERA experiment 
\footnote{This paper has been published in \texttt{\href{http://dx.doi.org/10.1103/PhysRevLett.115.121802}{Phys. Rev Lett. 115 121802 (2015)}
}}}

\author{N.~Agafonova}\affiliation{INR - Institute for Nuclear Research of the Russian Academy of Sciences, RUS-117312 Moscow, Russia}
\author{A.~Aleksandrov}\affiliation{INFN Sezione di Napoli, 80125 Napoli, Italy}
\author{A.~Anokhina}\affiliation{ SINP MSU - Skobeltsyn Institute of Nuclear Physics, Lomonosov Moscow State University, RUS-119991 Moscow, Russia}
\author{S.~Aoki}\affiliation{ Kobe University, J-657-8501 Kobe, Japan}
\author{A.~Ariga}\affiliation{Albert Einstein Center for Fundamental Physics, Laboratory for High Energy Physics (LHEP), University of Bern, CH-3012 Bern, Switzerland }
\author{T.~Ariga}\affiliation{Albert Einstein Center for Fundamental Physics, Laboratory for High Energy Physics (LHEP), University of Bern, CH-3012 Bern, Switzerland }
\author{D.~Bender}\affiliation{METU - Middle East Technical University, TR-06531 Ankara, Turkey}
\author{A.~Bertolin}\affiliation{INFN Sezione di Padova, I-35131 Padova, Italy}
\author{I.~Bodnarchuk}\affiliation{JINR - Joint Institute for Nuclear Research, RUS-141980 Dubna, Russia}
\author{C.~Bozza}\affiliation{Dipartimento di Fisica dell'Universit\`a di Salerno and ``Gruppo Collegato'' INFN, I-84084 Fisciano (Salerno), Italy}
% ...
\collaboration{The OPERA Collaboration} \noaffiliation \skip 0.25cm
%\thanks{thank denemeAAAAAAAAAAAAAAAAAAAAAAAAAAAAAAAAAAAAAAAAAAAAAAAAAAAAAAAAAAAAAAAAAAAAAAAAAAAAAAAAAAAAAAAAAAa dedicate }

\date{\today}
%\centerline{Published in Phys. Rev Lett. 115 121802 (2015)}
%\date{Published in Phys. Rev Lett. 115 121802 (2015)}

\begin{abstract}
The OPERA experiment was designed to search for $\nu_{\mu} \rightarrow \nu_{\tau}$ oscillations in appearance mode, i.e. by detecting the $\tau$ leptons produced in charged current $\nu_{\tau}$ interactions. The experiment took data from 2008 to 2012 in the CERN Neutrinos to Gran Sasso beam. The observation of the $\nu_{\mu} \rightarrow \nu_{\tau}$ appearance, achieved with four candidate events in a subsample of the data, was previously reported. In this Letter, a fifth $\nu_{\tau}$ candidate event, found in an enlarged data sample, is described. Together with a further reduction of the expected background, the candidate events detected so far allow us to assess the discovery of $\nu_{\mu}\rightarrow \nu_{\tau}$ oscillations in appearance mode with a significance larger than 5 $\sigma$.
% This paper has been published in \texttt{\href{http://dx.doi.org/10.1103/PhysRevLett.115.121802}{Phys. Rev Lett. 115 121802 (2015).}}
\end{abstract}

\maketitle

\emph{\label{sec:intor}Introduction.}$-$
Neutrino flavor transitions due to quantum mechanical mixing between neutrino flavors ($\nu_{e}$, $\nu_{\mu}$, $\nu_{\tau}$) and mass eigenstates ($\nu_{1}$, $\nu_{2}$, $\nu_{3}$) were proposed more than 50 years ago~\cite{MNS, pontecorvo2}. Several experiments on solar, atmospheric, reactor, and accelerator neutrinos have contributed to the understanding of these transitions, referred to as ``neutrino oscillations''~\cite{skamiokande1,skamiokande2,skamiokande3,sno,soudan,macro,k2k,kamland, minos}. In the atmospheric sector, the strong deficit of muon neutrinos observed by the Super-Kamiokande experiment in 1998 was the first compelling observation of neutrino oscillations~\cite{skamiokande1,skamiokande2,skamiokande3}. This result was later confirmed by the K2K~\cite{k2k} and MINOS experiments~\cite{minos}. However, for an unambiguous confirmation of three-flavor neutrino oscillations in the atmospheric sector, the detection of oscillated neutrinos in appearance mode was required. 

The OPERA experiment has been designed to search for $\nu_{\mu} \rightarrow \nu_{\tau}$ oscillations in appearance mode through the detection of the $\tau$ lepton produced in the $\nu_{\tau}$ charged current (CC) interactions. It has operated under low background conditions and with a signal-to-noise ratio as large as about 10. In 2010, a first $\nu_{\tau}$ candidate event was observed~\cite{opera_first}. In 2013, the Super-Kamiokande experiment reported evidence for $\nu_{\tau}$ appearance in the atmospheric $\nu_{\mu}$ flux with a signal-to-noise ratio of about one tenth~\cite{skamiokande_tau_appearance}. Since 2013, the detection by the OPERA experiment of three more candidate events reported in Refs.~\cite{opera_2nd_candidate,opera_3rd_candidate,opera_4th_candidate} has allowed us to claim the first observation of $\nu_{\mu} \rightarrow \nu_{\tau}$ oscillations in appearance mode with a 4.2 $\sigma$ significance~\cite{opera_4th_candidate}. In 2014, flavor transition with high purity in appearance mode has also been observed by the T2K experiment in the $\nu_{\mu} \rightarrow \nu_e$ channel~\cite{t2k}.

In this Letter, the observation of an additional $\nu_{\tau}$ candidate found in an enlarged data sample is reported. The significance of the $\nu_{\tau}$ appearance is updated  taking into account the new observed event and improvements in the background evaluation.

\emph{Neutrino beam, detector, and data sample.}$-$
\label{sec:dataSample}
The OPERA detector at the LNGS underground laboratory has been exposed from 2008 to 2012 to the CERN neutrinos to Gran Sasso (CNGS) $\nu_{\mu}$ beam~\cite{cngs}. A total exposure corresponding to 17.97~$\times$~10$^{19}$ protons on target (POT) resulted in 19 505 neutrino interactions in the target fiducial volume.

The topology of the neutrino interactions is recorded in emulsion cloud chamber detectors (ECC bricks) with submicrometric spatial resolution. Each brick is a stack of 56 1 mm thick lead plates, and 57 nuclear emulsion films with a 12.7 $\times$ 10.2 cm$^{2}$ cross section, a thickness of 7.5 cm corresponding to about 10 radiation lengths and a mass of 8.3 kg. 
In the bricks, the momenta of charged particles are measured by their multiple Coulomb scattering in the lead plates~\cite{opera_mcs}.
A changeable sheet (CS) doublet consisting of a pair of emulsion films~\cite{cs} is attached to the downstream face of each brick. The full OPERA target is segmented in about 150 000 bricks arranged in two identical supermodules (SMs). In each SM, the target section is made of 31 walls of ECC bricks. Downstream of each target wall, two orthogonal planes of electronic target trackers (TTs), made of 2.6 cm wide scintillator strips, record the position and deposited energy of charged particles~\cite{tt}.
 A spectrometer, consisting of iron core magnets instrumented with resistive plate chambers (RPCs) and drift tubes (precision tracker), is mounted downstream of each target module. The spectrometers are used to identify muons, determine their charge, and measure their momentum with an accuracy of about 20\%. A detailed description of the OPERA detector can be found in Ref.~\cite{detector_paper}.
 
A three-dimensional track in the electronic detector is tagged as a muon if the product of its length by the density along its path is larger than 660 g/cm$^{2}$~\cite{opera_ed}. An event is classified as $1\mu$ either if it contains at least one track tagged as a muon or if the total number of fired TT and RPC planes is larger than 19. The complementary sample is defined as $0\mu$. A muon track can be confirmed or discarded by measuring its trajectory all along the downstream bricks. The momentum-range correlation, the energy  loss near the stopping point and, eventually, the tagging of interaction or decay topologies may  contribute to assessing the muonic nature of the track beyond the electronic detector performance.

The analysis described below is extended to all $0\mu$ events and to $1\mu$ events with a muon momentum below 15~GeV/$c$ to reduce the background. The procedure starts with the use of the TT hits pattern to select the bricks possibly containing the neutrino interaction~\cite{opera_brick_finding}. These bricks are ordered according to their decreasing probability to contain the neutrino interaction vertex. The most probable brick (first brick hereafter) is then extracted from the target. If the neutrino interaction vertex is not found in this brick, it is searched for in the next brick in the probability ranking (second brick hereafter). Once the vertex has been located in a brick, a surrounding volume of about 2 cm$^{3}$ is scanned to detect $\tau$ leptons or other short-lived particle decays~\cite{charmpaper}. The details of the event analysis procedure are described in Ref.~\cite{opera_2nd_candidate}.

In this Letter, we report the analysis performed on the first and second bricks of all of the events recorded by OPERA. The event sample is about 15$\%$ larger than the one reported in Ref.~\cite{opera_4th_candidate}. 
The numbers of fully analysed events are given in Table~\ref{table:statistics} for each year of data taking.

\begin{table}
\centering
\begin{tabular}{lcccccc} \hline\hline
                              &2008 &2009 &2010 &2011 &2012 &Total \\ \hline
POT (10$^{19}$)             &1.74 &3.53 &4.09 &4.75 &3.86 &17.97 \\
0$\mu$ events                 &149  &253  &268 &270 &204 &1144 \\
1$\mu$ events ($p_{\mu}<$15 GeV/$c$) &542 &1020 &968  &966 &768 &4264 \\
Total events                  &691 &1273 &1236 &1236 &972 &5408 \\
Detected $\nu_{\tau}$ candidates & &1 & &1 &3 &5\\ \hline
\hline
\end{tabular}
\caption{Number of events used in this analysis and the detected $\nu_{\tau}$ candidates for each run year.}
\label{table:statistics}
\end{table}

\begin{figure*}[ht!]
\includegraphics[width=0.75\textwidth]{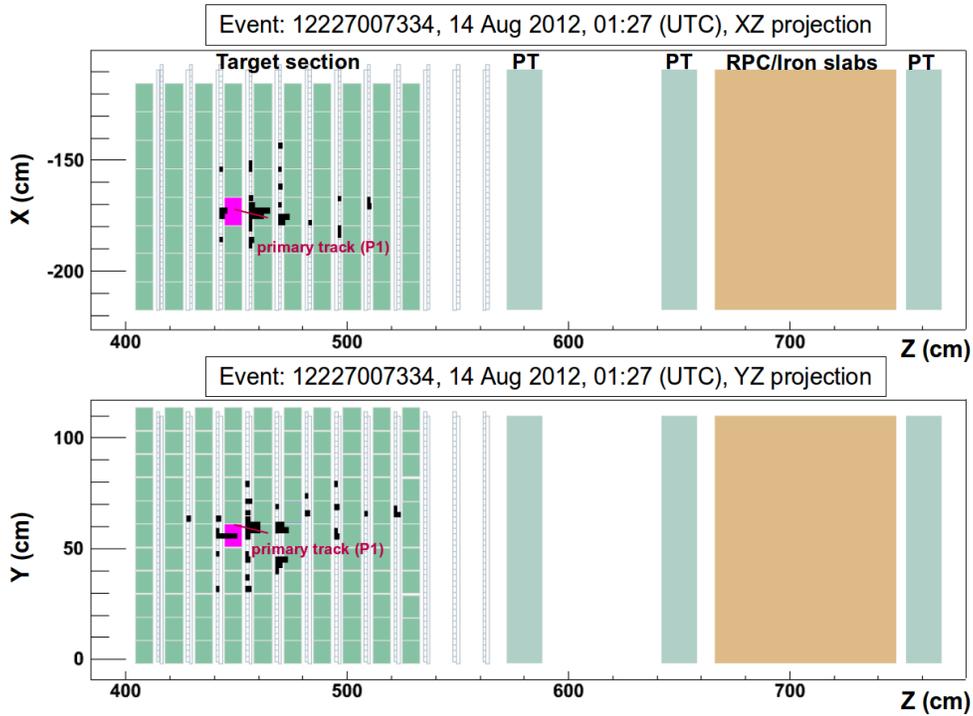}
\caption{Display of the $\nu_{\tau}$ candidate event as seen by the electronic detectors in the $x$-$z$ projection (top panel) and $y$-$z$ projection (bottom panel). The OPERA (right-handed) reference frame is oriented such that the $y$ axis is perpendicular to the hall floor and pointing up; the $z$ axis is orthogonal to the brick walls and is oriented as the incoming neutrinos. The angle between the neutrino direction and the $z$ axis projected into the $yz$ plane is 58 mrad. The brick containing the neutrino interaction is highlighted in magenta. The solid line shows the direction of the primary track P1 (see the text) at its most upstream point as reconstructed in the emulsion detectors.}
\label{fig:5ev_TT_display}
\end{figure*}

\emph{The new $\nu_{\tau}$ candidate event.}$-$
\label{sec:newCandidate}
The new $\nu_{\tau}$ candidate event reported here occurred on August 14, 2012 in the second SM, seven brick walls upstream of the spectrometer.
As shown in Fig.~\ref{fig:5ev_TT_display}, the activity in the TT is limited to the six walls downstream of the vertex brick. The event is classified as $0\mu$. The visible energy of the event is \mbox{$12\pm4$} ~GeV.

A converging pattern of tracks in the CS hints to a possible vertex in the brick. Following these tracks inside the brick, the neutrino interaction vertex (the primary vertex) was localised in the 42$^{nd}$ lead plate from the downstream face of the brick. 

The primary vertex consists of the $\tau$ candidate track, which exhibits a kink topology, and a charged particle track (P1). 
The distance of closest approach between the $\tau$ candidate and P1 is 0.1 $\mu$m, compatible with zero within the tracking resolution. In addition to the $\tau$ lepton and P1, four forward-going and two backward-going nuclear fragments pointing to the primary vertex are observed. 

\begin{figure}[ht!]
  \centering
   \includegraphics[width=0.45\textwidth]{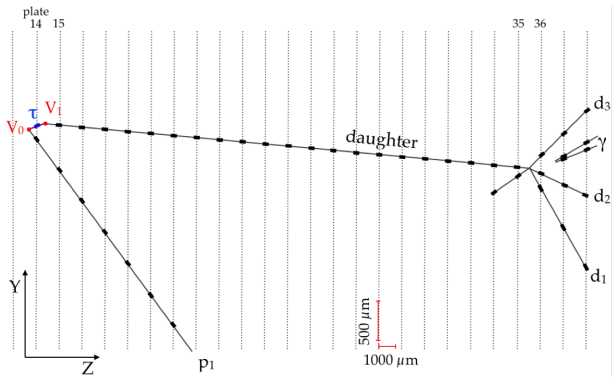}
\caption{Event display of the fifth $\nu_{\tau}$ candidate event in the horizontal projection longitudinal to the neutrino direction. The primary and secondary vertices are indicated as $V_{0}$ and $V_{1}$, respectively. The black stubs represent the track segments as measured in the films.
 }
\label{fig:5th_topology_display}
\end{figure}

The $\tau$ candidate decays at a flight length of \mbox{$960\pm30$}$\mu$m into one charged particle which interacts after crossing 22 plates and can thus be unambiguously identified as a hadron. The interaction of the daughter particle produces four charged particles and a photon. Figure~\ref{fig:5th_topology_display} shows the display of the event as reconstructed in the brick.

The difference in angle between the $\tau$ candidate track and the daughter particle track, $\theta_{\mathrm{kink}}$, is \mbox{$90\pm2$} ~mrad. The daughter track has an impact parameter of
\mbox{$83\pm5$}$~\mu$m with respect to the primary vertex.
The $z$ coordinate of the decay vertex, $z_{\mathrm{dec}}$, measured from the downstream face of the lead plate containing the primary vertex, is \mbox{$630\pm30$} $\mu$m. 
A search for nuclear fragments has been performed both upstream and downstream of the kink vertex up to $\tan \theta=3$~\cite{fukuda_theta3} (with $\theta$ being the angle of the track with respect to the $z$ axis). No fragment is found. This result strongly reduces the probability of the secondary vertex being due to hadronic interaction.

The charged particle producing the primary track (P1) has a measured momentum of \mbox{$1.0\pm0.1$} GeV/$c$. It is identified as a hadron from its interaction in the downstream brick. This, together with the negative search for large angle tracks~\cite{fukuda_las}, allows us to rule out the presence of a muon at the primary vertex (expected for $\nu_{\mu}$CC related backgrounds). The linear density of grains along the track left by a particle is correlated with the energy loss of the particle. The ratio between the grain density of track P1 and that of the $\tau$ daughter track is \mbox{$1.45\pm0.06$}, to be compared with the \mbox{$1.38\pm0.14$} expected for a proton to minimum ionizing particle ratio. Therefore, track P1 is most likely left by a proton~\cite{fukuda_grain}.

A search for photon conversions possibly pointing to the primary and secondary vertices was performed. None was found.

\begin{figure}[ht!]
  \centering
  \includegraphics[width=.45\textwidth]{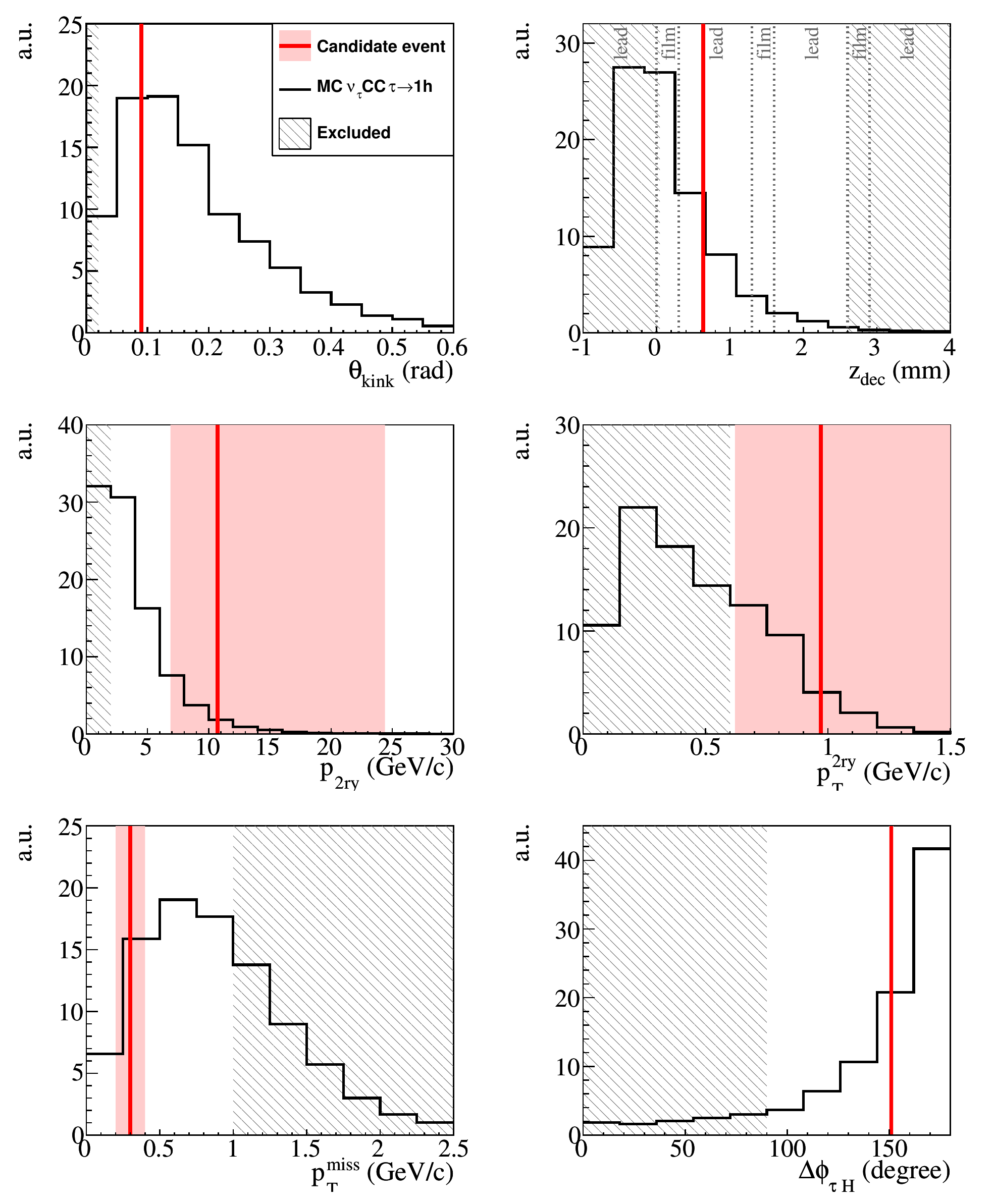}
  \caption{Monte Carlo distributions of the kinematical variables for $\nu_{\tau}$ events passing all the location and decay search chain with $\tau \rightarrow 1h$ decay topology. Red lines show the measured values for the candidate event and the corresponding errors. Grey areas show the regions excluded by the selection criteria.}
  \label{fig:5th_kin_histos}
\end{figure}

The scalar sum of the momenta of all particles measured in the brick, $p_{\mathrm{sum}}$, is $12^{+14}_{-4}$ GeV/$c$. The measured values of the kinematical parameters and the corresponding predefined selection criteria are summarised in Table.~\ref{table:kinematical_values}. In the table, $p^{2ry}$ and $p_{T}^{2ry}$ are the momentum and the transverse momentum of the decay daughter, respectively, $p_{T}^{\mathrm{miss}}$ is the missing transverse momentum at the primary vertex and $\Delta \phi_{\tau H}$ is the angle between the $\tau$ candidate direction and the hadron direction in the plane transverse to the beam direction.
The measured values of the kinematical parameters of the candidate event satisfy all of the selection criteria for the $\tau \rightarrow 1h$ channel.
The Monte Carlo distributions of the variables and the measured values are shown in Fig.~\ref{fig:5th_kin_histos}. 

\begin{table}
  \centering
    \begin{tabular}{cccc}
    \hline\hline
    Parameter & Measured value & Selection criteria \\
    \hline
    $\Delta \phi_{\tau H}$ ($^{\mathrm{o}}$) & $151 \pm 1$ & $>$ 90 \\
    $p_{T}^{\mathrm{miss}}$ (GeV/$c$) &$0.3 \pm 0.1$ & $<$ 1 \\
    $\theta_{\mathrm{kink}}$ (mrad) & $90 \pm 2$ & $>$ 20 \\
    $z_{\mathrm{dec}}$ ($\mu m$) & $630 \pm 30$ & $[44,2600]$ \\
    $p^{2ry}$ (GeV/$c$) & $11^{+14}_{-4}$ & $>$ 2 \\
    $p_{T}^{2ry}$ (GeV/$c$)& $1.0^{+1.2}_{-0.4}$ & $>$ 0.6 (no $\gamma$ attached) \\  \hline \hline
  \end{tabular}
  \caption{Kinematical parameters considered for the  $\tau \rightarrow 1h$ decay channel selection: measured values for the new candidate event and predefined cuts are reported in the second and third columns, respectively}.
  \label{table:kinematical_values}
\end{table}

\emph{Signal and background estimation.}$-$
\label{sec:significanceDm2}
The expected numbers of signal and background events as well as the number of detected $\nu_{\tau}$ candidates for each decay channel are summarised in Table~\ref{table:expected_signal_background}. Assuming $\Delta m^{2}_{23} = 2.44 \times 10^{-3}$ eV$^{2}$~\cite{pdg} and $\sin^{2}2\theta_{23} = 1$, the total expected signal is \mbox{$2.64\pm0.53$} events, whereas the total background expectation is \mbox{$0.25\pm0.05$} events.

The numbers of expected signal and background events are estimated from the simulated CNGS flux~\cite{fluka_cngs}. The expected detectable signal events in the $0\mu$ and $1\mu$ samples are obtained using the reconstruction efficiencies and the $\nu_\tau$ event rate in the flux normalised to the detected $\nu_{\mu}$ interactions. A similar normalisation procedure is also used in the background expectation. The details of the signal and background estimation are described in Ref.~\cite{opera_2nd_candidate}.

The systematic uncertainty associated with the signal takes into account contributions from the limited knowledge of the $\nu_{\tau}$ cross section and uncertainties on the signal detection efficiency. For the signal central value, the default implementation for the $\nu_{\tau}$ cross-section contained in the GENIE v2.6 simulation program is used~\cite{genie}. A 10\% model-related systematic uncertainty can be estimated by considering the maximal deviations from the central value of the expected number of $\nu_\tau$ candidates obtained when considering all of the available theoretical predictions. The only existing measurement of the $\nu_{\tau}$ cross section is a very low-statistics one by the DONUT experiment~\cite{donut_cross_section}. Owing to the fact that the $\nu_\tau$ signal expectation is calculated by using location efficiencies determined from the $1\mu$ and $0\mu$ data samples, this value is at first order insensitive to systematic effects on efficiencies up to the primary vertex location level. Further confidence on the global efficiency estimation is obtained by considering the charm data sample for which good agreement is found between the 50 observed events and the expectation (\mbox{$54\pm4$}) provided by the neutrino-induced charm production cross section and the detector simulation~\cite{chorus_charm,charmpaper}. Additional uncertainties on the number of expected $\nu_{\tau}$ candidates arise from the experimental knowledge of $\theta_{23}$ and $\Delta m_{23}^{2}$ (10 \%), and from the uncertainty in the efficiency for tagging $\tau$ lepton decays (15\%). The latter contribution arises from the statistical error of the sample of $\nu_{\mu}^{\mathrm{CC}}$ events with charm production which was used for validation. The CNGS flux uncertainty plays a minor role since the expected number of $\nu_{\tau}$ events is determined from the detected $\nu_{\mu}$ interactions used as a normalisation sample. The simulation of the kinematical properties of the final state was performed using the NEGN generator~\cite{dario}, which takes the polarisation of $\tau$ leptons into account ($\tau$ decay library TAUOLA~\cite{taulo}). The associated systematic uncertainty on the expected number of $\tau$ decays in all channels is estimated at the level of a few percent~\cite{tauPolarization}. The total systematic uncertainty on the expected signal is then set to 20\%.

The main processes contributing to the background for the $\nu_\tau$ appearance search are charmed particle decays, hadronic interactions and large-angle muon scattering (LAS). The corresponding contributions are estimated by simulation studies validated with real data samples. Using the measured sample of CNGS $\nu_{\mu}$CC interactions with charm production, the uncertainty on the charm background has been estimated to about 20\%~\cite{charmpaper}. This includes a contribution from the experimental uncertainty on the charm cross section (8\%~\cite{chorus_charm}), the hadronisation fraction (10\%), and the statistical error of the CNGS charm control sample (15\%). Hadronic background has an estimated uncertainty of 30\% from data-driven measurements of test-beam pion interactions in the OPERA bricks~\cite{tohopaper}.

With respect to what was reported in Ref.~\cite{opera_2nd_candidate}, an additional improvement in the estimation of the LAS background in the $\tau \rightarrow \mu$ decay channel has been achieved~\cite{las_reduction_andrea}. The LAS rate is estimated using a GEANT4-based simulation implementing a mixed-approach algorithm with $ad \ hoc$ modifications to take into account the effect of the nuclear form factor at the involved transferred momenta (of the order of a few fm$^{-1}$). A Saxon-Woods charge density is assumed with parameters derived from fits to data. Scattering off individual protons is also taken into account. The simulation is benchmarked on experimental data including scattering of 2 GeV/c muons on a 12.6 mm lead target, 7.3 GeV/$c$ and 11.7 GeV/$c$ muons on a 14.4 mm thick copper target and 0.512 GeV/$c$ electrons on a 0.217 mm lead target~\cite{las_reduction_1,las_reduction_2,las_reduction_3}. From this study, it follows that the number of LAS background events that satisfy the $\tau \rightarrow \mu$ selection criteria amounts to [\mbox{$1.2\pm0.1(\rm{stat.})\pm0.6 (\rm{sys.})$}] $\times$ 10$^{-7}$/$\nu_{\mu}^{\mathrm{CC}}$ interactions, well below the conservative value considered in our past publications.

\begin{table*}
\centering
  \resizebox{0.85\textwidth}{!}
            {%
              \begin{tabular}{c|cccc|c|c}
                \hline\hline
                \multirow{ 2}{*}{Channel} &\multicolumn{4}{c|}{Expected background} & \multirow{ 2}{*}{Expected signal} & \multirow{ 2}{*}{Observed} \\ \cline{2-5}
                &Charm &Had. reinterac. &Large $\mu$ scat. &Total & & \\\hline
                $\tau \rightarrow 1h$ 	&$0.017 \pm 0.003$  &$0.022 \pm 0.006$ 	&  &$0.04 \pm 0.01$ &$0.52 \pm 0.10$ &3 \\ 
                $\tau \rightarrow 3h$ 	&$0.17 \pm 0.03$ 	&$0.003 \pm 0.001$ 	&  &$0.17 \pm 0.03$ &$0.73 \pm 0.14$ &1\\ 
                $\tau \rightarrow \mu$ 	&$0.004 \pm 0.001$ 	& 	            &$0.0002 \pm 0.0001$  &$0.004 \pm 0.001$ &$0.61 \pm 0.12$ &1\\
                $\tau \rightarrow e$ 	&$0.03 \pm 0.01$ 	&            	&  &$0.03 \pm 0.01$ &$0.78 \pm 0.16$ &0\\
                \hline
                Total  &$0.22 \pm 0.04$ &$0.02 \pm 0.01$ &$0.0002 \pm 0.0001$ &$0.25 \pm 0.05$ &$2.64 \pm 0.53$ &5\\ \hline\hline
              \end{tabular} 
            }
            \caption{Expected signal and background events for the analysed data sample.}
            \label{table:expected_signal_background}
\end{table*}

\emph{Results.}$-$
In this analysis, the observed number of $\nu_\tau$ candidates $n_i$ for each individual $\tau$ decay channel $i$ is considered as an independent Poisson process with expectation $\mu s_i + b_i$. The expected signal and background events, $s_i$ and $b_i$ respectively, are taken from Table~\ref{table:expected_signal_background}; the signal strength factor $\mu$ is a continuous multiplicative parameter for the expected signal. The background-only hypothesis corresponds to $\mu$ = 0, and the nominal signal to $\mu$ = 1.
 
The significance of the observed $\nu_\tau$ candidates is evaluated as the probability  that the background can produce a fluctuation greater than or equal to the observed data. Two test statistics are used for the computation; in both cases, the test statistics values of the observed data are compared with sampling distributions obtained with pseudoexperiments.
 
The first test statistics is based on the Fisher's method. For the background-only hypothesis (i.e., $\mu=0$), the p values $p_i$ of each individual channel  (calculated as the integral of the Poisson distribution for values larger or equal to the observed number of candidates) are combined into an estimator $p^\star= \prod_i p_i$~\cite{pstar,pstar_cdf}. By comparing the observed $p^\star_{\mathrm{data}}$ with the sampling distribution of $p^\star$, a (one-side) significance of 5.1 standard deviations is obtained, corresponding to a background fluctuation probability of \mbox{$1.1\times10^{-7}$}.

The second test statistics is based on the one-sided profile likelihood ratio $\lambda(\mu)$~\cite{pdg}. This test statistic is used to quantify the discrepancy between the data and a certain hypothesised value of $\mu$. The significance, the level of disagreement between the observed data and the $\mu = 0$ hypothesis, is computed by comparing $\lambda_{\mathrm{data}}(\mu = 0)$ with the corresponding sampling distribution of $\lambda(\mu = 0)$. The likelihood, which includes Gaussian terms to account for the background uncertainties, is

\begin{equation}
 \mathcal{L} = \prod_{i=1}^{4} \mathrm{Poisson}(n_{i}|\mu s_{i}+\beta_{i})\mathrm{Gauss}(\beta_{i}|b_{i},\sigma_{b_i}),
\end{equation}
where $\sigma_{b_i}$ is the background uncertainty for channel $i$ (from Table~\ref{table:expected_signal_background}) and $\beta_i$ are the background parameters Gaussian modelled. Two different implementations of the method, one based on a custom code and the other one based on RooStats~\cite{roostat}, have been used with both giving a significance of 5.1 standard deviations. 

A simple compatibility test of the observed data with the expectations
from the  neutrino oscillation hypothesis ($\mu=1$) is given by the
best-fit signal strength at 90\% C.L., $\hat{\mu} = 1.8^{+1.8}_{-1.1},$ which is
consistent with unity. Another test was made  by performing
pseudoexperiments to sample the distribution of the data assuming $\mu=1$
and taking into account the uncertainties on the expected signal and
background. The probability of data being less likely or equal to the
observed ones is 6.4\%. If we consider the total number of $\nu_{\tau}$ candidates regardless of the distribution into decay channels, the probability of observing five or more candidates with an expectation of 2.64 signal plus 0.25 background events is 17\% from Poisson statistics.

The 90\% confidence interval for $\Delta m_{23}^{2}$ has been estimated with three different approaches using the profile likelihood ratio, the Feldman-Cousins method, and Bayesian statistics.  Assuming full mixing, the best fit is $\Delta m^{2}_{23} =$ 3.3 $\times$ 10$^{-3}$ eV$^{2}$ with a 90\% C.L. interval of \big[2.0, 5.0\big] $\times$ 10$^{-3}$ eV$^{2}$, the differences among the three methods being negligible.

\emph{Conclusions.}$-$
\label{sec:conclusion}
This Letter reports the analysis of a data sample including the first and the second most probable bricks for all runs, with a corresponding increase of the statistics of about 15\% with respect to Ref.~\cite{opera_4th_candidate}. In this enlarged data sample, a fifth $\tau$ neutrino candidate has been found. Furthermore, a revision of the background estimate in the muonic decay channel has been performed. Given the low background level and the observed number of $\nu_{\tau}$ candidate events, we report the discovery of a $\nu_\tau$ appearance in the CNGS neutrino beam with a significance of 5.1~$\sigma$.

\begin{acknowledgments}
We acknowledge CERN for the successful operation of the CNGS facility and INFN for the continuous support given to the experiment through its LNGS laboratory. We acknowledge funding from our national agencies: Fonds de la Recherche Scientifique-FNRS and Institut Inter Universitaire des Sciences Nucleaires for Belgium; MoSES for Croatia; CNRS and IN2P3 for France; BMBF for Germany; INFN for Italy; JSPS, MEXT, the QFPU-Global COE program of Nagoya University, and Promotion and Mutual Aid Corporation for Private Schools of Japan for Japan; SNF, the University of Bern and ETH Zurich for Switzerland; the Russian Foundation for Basic Research (Grant No. 12-02-12142 ofim), the Programs of the Presidium of the Russian Academy of Sciences (Neutrino physics and Experimental and Theoretical Researches of Fundamental Interactions), and the Ministry of Education and Science of the Russian Federation for Russia; the National Research Foundation of Korea (Grant No. NRF-2013R1A1A2061654) for Korea; and TUBITAK, the Scientific and Technological Research Council of Turkey for Turkey. We thank the IN2P3 Computing Centre (CC-IN2P3) for providing computing resources.
\end{acknowledgments}


\begin{thebibliography}{99}
\bibitem{MNS}
	Z. Maki, M. Nakagawa, and S. Sakata, \texttt{\href{http://dx.doi.org/10.1143/PTP.28.870}{Progr. Theor. Phys. \textbf{28}, 870 (1962).}}

\bibitem{pontecorvo2}
  B. Pontecorvo, Zh. Eksp. Teor. Fiz. \textbf{53}, 1717 (1967).

\bibitem{skamiokande1}
	Y. Fukuda {\sl et al}. (Super-Kamiokande Collaboration), \texttt{\href{http://dx.doi.org/10.1103/PhysRevLett.81.1562}{Phys. Rev. Lett. \textbf{81}, 1562 (1998).}}
    
\bibitem{skamiokande2}
	K. Abe {\sl et al}. (Super-Kamiokande Collaboration), \texttt{\href{http://dx.doi.org/10.1103/PhysRevLett.97.171801}{Phys. Rev. Lett. \textbf{97}, 171801 (2006).}}

\bibitem{skamiokande3}
	R. Wendell {\sl et al}. (Super-Kamiokande Collaboration), \texttt{\href{http://dx.doi.org/10.1103/PhysRevD.81.092004}{Phys. Rev. D \textbf{81}, 092004 (2010).}}
    
\bibitem{sno}
	Q. R. Ahmad {\sl et al}. (SNO Collaboration), \texttt{\href{http://dx.doi.org/10.1103/PhysRevLett.87.071301}{Phys. Rev. Lett. \textbf{87}, 071301 (2001).}}

\bibitem{soudan}
	W. W. M. Allison {\sl et al}. (Soudan-2 Collaboration), \texttt{\href{http://dx.doi.org/10.1103/PhysRevD.72.052005}{Phys. Rev. D \textbf{72}, 052005 (2005).}}

\bibitem{macro}
	M. Ambrosio {\sl et al}. (MACRO Collaboration), \texttt{\href{http://dx.doi.org/10.1140/epjc/s2004-01951-9}{Eur. Phys. J. C. \textbf{36}, 323 (2004).}}

\bibitem{k2k}
	M. H. Ahn {\sl et al}. (K2K Collaboration), \texttt{\href{http://dx.doi.org/10.1103/PhysRevD.74.072003}{Phys. Rev. D \textbf{74}, 072003 (2006).}}

\bibitem{kamland}
	S. Abe {\sl et al}. (KamLAND Collaboration), \texttt{\href{http://dx.doi.org/10.1103/PhysRevLett.100.221803}{Phys. Rev. Lett. \textbf{100}, 221803 (2008).}}

\bibitem{minos}
	P. Adamson {\sl et al}. (MINOS Collaboration), \texttt{\href{http://dx.doi.org/10.1103/PhysRevLett.106.181801}{Phys. Rev. Lett. \textbf{106}, 181801 (2011).}}

\bibitem{opera_first}
	N. Agafonova {\sl et al}. (OPERA Collaboration), \texttt{\href{http://dx.doi.org/10.1016/j.physletb.2010.06.022}{Phys. Lett. B \textbf{691}, 138 (2010).}}

\bibitem{skamiokande_tau_appearance}
	K. Abe {\sl et al}. (Super-Kamiokande Collaboration), \texttt{\href{http://dx.doi.org/10.1103/PhysRevLett.110.181802}{Phys. Rev. Lett. \textbf{110}, 181802 (2013).}}

\bibitem{opera_2nd_candidate}
	N. Agafonova {\sl et al}. (OPERA Collaboration), \texttt{\href{http://dx.doi.org/10.1007/JHEP11(2013)036}{JHEP \textbf{11}, 036 (2013).}}

\bibitem{opera_3rd_candidate}
	N. Agafonova {\sl et al}. (OPERA Collaboration), \texttt{\href{http://dx.doi.org/10.1103/PhysRevD.89.051102}{Phys. Rev. D \textbf{89}, 051102(R) (2014).}}

\bibitem{opera_4th_candidate}
	N. Agafonova {\sl et al}. (OPERA Collaboration), \texttt{\href{http://dx.doi.org/10.1093/ptep/ptu132}{Prog. Theor. Exp. Phys. \textbf{2014}, 101C01 (2014).}}

\bibitem{t2k}
	K. Abe {\sl et al}. (T2K Collaboration), \texttt{\href{http://dx.doi.org/10.1103/PhysRevLett.112.061802}{Phys. Rev. Lett. \textbf{112}, 061802 (2014).}}

\bibitem{cngs}
  K. Elsener, Reports No. CERN 98-02 and No. INFN/AE-98/05, 1998; R. Bailey {\sl et al}., Reports No. CERN-SL-99-034-DI and No. INFN-AE-99-05, 1999, addendum to Reports No. CERN 98-02 and No. INFN-AE-98-05, 1998; CNGS Web site, \texttt{\href{http://proj-cngs.web.cern.ch/proj-cngs}{http://proj-cngs.web.cern.ch/proj-cngs.}}

\bibitem{opera_mcs} 
	N. Agafonova {\sl et al}. (OPERA Collaboration), \texttt{\href{http://dx.doi.org/10.1088/1367-2630/14/1/013026}{New J. Phys. \textbf{14}, 013026 (2012).}}

\bibitem{cs}
	A. Anokhina {\sl et al}. (OPERA Collaboration) \texttt{\href{http://dx.doi.org/10.1088/1748-0221/3/07/P07005}{JINST \textbf{3}, P07005 (2008).}}
    
\bibitem{tt}
	T. Adam {\sl et al}. (OPERA Collaboration), \texttt{\href{http://dx.doi.org/10.1016/j.nima.2007.04.147}{Nucl. Instrum. Methods A \textbf{577}, 523 (2007).}}

\bibitem{detector_paper}
	R. Acquafredda {\sl et al}. (OPERA Collaboration), \texttt{\href{http://dx.doi.org/10.1088/1748-0221/4/04/P04018}{JINST \textbf{4}, P04018 (2009).}}
      
\bibitem{opera_ed}
  N. Agafonova {\sl et al}. (OPERA Collaboration), \texttt{\href{http://dx.doi.org/10.1088/1367-2630/13/5/053051}{New J. Phys. \textbf{13}, 053051 (2011).}}

\bibitem{opera_brick_finding}
	Y. A. Gornushkin, S. G. Dmitrievsky and A. V. Chukanov, \texttt{\href{http://dx.doi.org/10.1134/S1547477115010100}{Phys.Part.Nucl.Lett. \textbf{12}, 89 (2015).}}

\bibitem{charmpaper}
N. Agafonova {\sl et al}. (OPERA Collaboration), \texttt{\href{http://dx.doi.org/10.1140/epjc/s10052-014-2986-0}{Eur.Phys.J. C\textbf{74}, 2986 (2014).}}

\bibitem{fukuda_theta3}
  T. Fukuda, S. Fukunaga, H. Ishida, K. Kodama, T. Matsuo, S. Mikado, S. Ogawa, H. Shibuya, and J. Sudo, \texttt{\href{http://dx.doi.org/10.1088/1748-0221/8/01/P01023}{JINST \textbf{8}, P01023 (2013).}}

\bibitem{fukuda_las}
  T. Fukuda {\sl et al}., \texttt{\href{http://dx.doi.org/10.1088/1748-0221/9/12/P12017}{JINST \textbf{9} P12017 (2014).}}

\bibitem{fukuda_grain}
  T. Fukuda, \texttt{\href{http://operaweb.lngs.infn.it/Opera/publicnotes/OPERA_note_179.pdf}{OPERA Public Note No. 179, (2015).}}

\bibitem{pdg}
	K.A Olive {\sl et al}. (Particle Data Group) \texttt{\href{http://dx.doi.org/10.1088/1674-1137/38/9/090001}{Chin. Phys. C\textbf{38}, 090001 (2014).}}

\bibitem{fluka_cngs}
See \texttt{\href{http://www.mi.infn.it/$\sim$psala/Icarus/cngs.html}{http://www.mi.infn.it/$\sim$psala/Icarus/cngs.html}}

\bibitem{genie}
C. Andreopoulos {\sl et al}., \texttt{\href{http://dx.doi.org/10.1016/j.nima.2009.12.009}{Nucl. Instrum. Methods. A \textbf{614}, 87 (2010).}}

\bibitem{donut_cross_section}
K. Kodama {\sl et al}. (The DONUT Collaboration), \texttt{\href{http://dx.doi.org/10.1103/PhysRevD.78.052002}{Phys. Rev. D\textbf{78}, 05002 (2008).}}

\bibitem{chorus_charm}
  A. Kayis-Topaksu {\sl et al}. (CHORUS Collaboration), \texttt{\href{http://dx.doi.org/10.1088/1367-2630/13/9/093002}{New J. Phys. \textbf{13}, 093002 (2011).}}

\bibitem{dario}
  D. Autiero, \texttt{\href{http://dx.doi.org/10.1016/j.nuclphysbps.2004.11.168}{Nucl. Phys. B, Proc. Suppl. \textbf{139}, 253 (2005).}}

\bibitem{taulo}
  S. Jadach, Z. Wa̧s, R. Decker, and J.H. Kühn, \texttt{\href{http://dx.doi.org/10.1016/0010-4655(93)90061-G}{Comput. Phys. Commun., \textbf{76} 361 (1993)}}

\bibitem{tauPolarization}
  M. Aoki, K. Hagiwara, K. Mawatari, and H. Yokoya, \texttt{\href{http://dx.doi.org/10.1016/j.nuclphysb.2005.08.036}{Nucl.Phys.B \textbf{727}, 163 (2005).}}

\bibitem{tohopaper}
  H. Ishida {\sl et al}., \texttt{\href{http://dx.doi.org/10.1093/ptep/ptu119}{Prog. Theor. Exp. Phys., 093C01 (2014).}}

\bibitem{las_reduction_andrea}
	A.Longhin, A.Paoloni, and F.Pupilli, \texttt{\href{http://dx.doi.org/10.1109/TNS.2015.2473674}{IEEE Trans. Nucl. Sc. \textbf{62}, 2216 (2015).}}

\bibitem{las_reduction_1}
  B. Frois, J.B. Bellicard, J.M. Cavedon, M. Huet, P. Leconte, P. Ludeau, A. Nakada, Xuan Ho Phan, and I. Sick, \texttt{\href{http://dx.doi.org/10.1103/PhysRevLett.38.152}{Phys. Rev. Lett. \textbf{38}, 152 (1977).}}

\bibitem{las_reduction_2}
S.A. Akimenko, V.I. Belousov, A.M. Blik, G.I. Britvich, V.N. Kolosov, V.M. Kutin, B.N. Lebedev, V.N. Peleshko, Ya.N. Rastsvetalov, and A.S. Solovev, \texttt{\href{http://dx.doi.org/10.1016/0168-9002(86)90990-3}{Nucl. Instrum. Methods Phys. Res., Sec. A \textbf{243}, 518 (1986).}}

\bibitem{las_reduction_3}
G.E. Masek, L.D. Heggie, Y.B. Kim, and R.W. Williams, \texttt{\href{http://dx.doi.org/10.1103/PhysRev.122.937}{Phys. Rev. \textbf{122}, 937 (1961).}}
    
\bibitem{pstar}
	O. Sato, \texttt{\href{http://operaweb.lngs.infn.it/Opera/publicnotes/OPERA_note_173.pdf}{OPERA Public Note No. 173, (2014).}}

\bibitem{pstar_cdf}
%  L. Demortier, \texttt{\href{\tbox{http://www-cdf.fnal.gov/$^{\sim}$luc/statistics/cdf8662.pdf}}{http://www-cdf.fnal.gov/$^{\sim}$luc/statistics/cdf8662.pdf}}
L. Demortier, \texttt{\href{http://www-cdf.fnal.gov/\~luc/statistics/cdf8662.pdf}{http://www-cdf.fnal.gov/\~luc/ \ statistics/cdf8662.pdf}}

    
\bibitem{roostat}
	See \texttt{\href{https://twiki.cern.ch/twiki/bin/view/RooStats/WebHome}{https://twiki.cern.ch/twiki/bin/view/RooStats/\ WebHome}}
    
\end{thebibliography}
\end{document}